\documentclass[11pt]{article}
\usepackage{amssymb,amsmath}
\usepackage{graphicx}
\usepackage{hyperref}
\usepackage{cite}
\usepackage{subfigure}
\usepackage{epsf}

\hoffset= -2cm \voffset= -2cm \vspace{2cm}

\newcommand{\be}{\begin{equation}}
\newcommand{\ee}{\end{equation}}

\begin{document}
\title{Axial anomaly as a collective  effect of meson spectrum
}
\date{}
\author{Yaroslav~N.~Klopot$^1$\footnote{{\bf e-mail}: klopot@theor.jinr.ru}\;,\,
        Armen~G.~Oganesian$^2$\footnote{{\bf e-mail}: armen@itep.ru}\;\, and \
        Oleg~V.~Teryaev$^1$\footnote{{\bf e-mail}: teryaev@theor.jinr.ru}}
\maketitle
\begin{center}
{$^{1}$\em Bogoliubov Laboratory of Theoretical Physics,\\ Joint Institute for Nuclear Research,\;\; \\ Dubna 141980, Russia.\\
  $^{2}$Institute of Theoretical and Experimental Physics,\;\;\\B. Cheremushkinskaya 25, Moscow 117218, Russia}
\end{center}
\vspace{2cm}

\begin{abstract}
We study the transition form factors of the light mesons in the
kinematics, where one photon is real and other is virtual. Using
the dispersive approach to axial anomaly we show that the axial
anomaly in this case reveals itself as a collective effect of
meson spectrum. This allows us to get the relation between
possible corrections to continuum and to lower states within QCD
method which does not rely on factorization hypothesis. We show,
relying on the recent data of the BaBar Collaboration,  that the
relative correction to continuum is quite small, and small
correction to continuum can dramatically change the pion form
factor.
\end{abstract}

\section{Introduction}

The phenomenon of axial anomaly \cite{Bell:1969ts,Adler:1969gk} is known to be one of the most
subtle effects of quantum field theory. Perhaps  the most vivid
manifestation of it in particle physics can be found in
two-photon decays of pseudoscalars. Usually the differential form
of the axial anomaly is utilized to study this kind of processes.
However, less known dispersive approach to axial anomaly
(\cite{Dolgov:1971ri,Horejsi:1985qu}, for a review, see
\cite{Ioffe:2006ww}) leads to anomaly sum rule (ASR) relation,
which provides a very powerful tool for study of various
characteristics of meson spectrum. The absence of corrections to
ASR allows to get certain exact relations between characteristics
of hadrons, such as decay width \cite{Ioffe:2007eg} and relations
between mixing parameters of pseudoscalars
\cite{Klopot:2009cm,Klopot:2010sm}.

In the study of two-photon decays of pseudoscalars, usually  the
case of real photons is considered. However, ASR is valid for
virtual photons also \cite{Horejsi:1994aj,Veretin:1994dn} which
leads to several  interesting applications. As we will see, in
the case of one real and one virtual photon
\cite{Horejsi:1994aj}   the ASR takes into account the infinite
number of meson states (which can contribute to it due to their
quantum numbers), i.e. axial anomaly reveals itself as a
collective effect of meson spectrum. This nontrivial fact
together with the exactness of ASR gives us a tool to study
different characteristics of meson spectrum (form factors of
mesons, relation between possible corrections to lower meson
states and continuum).

The experimental measurements  of the photon-pion
$\gamma\gamma^*\to \pi^0$ transition form factor
$F_{\pi\gamma}(Q^2)$ in the range of photon virtualities $Q^2<9$
$GeV^2$ were performed by CELLO \cite{Behrend:1990sr} and CLEO
\cite{Gronberg:1997fj} Collaborations. The measured values of
$F_{\pi\gamma}(Q^2)$ appeared to be consistent with  predictions based on  factorization
approach to pQCD. Surprisingly, recent data of BaBar
Collaboration  on $F_{\pi\gamma}(Q^2)$ \cite{Aubert:2009mc}, which is available in the range   $4<Q^2<40$
$GeV^2$, have shown a strong disagreement  with the pQCD predicted behaviour of
$\gamma\gamma^*\to \pi^0$ transition form factor.

Though the BaBar data in the range $Q^2<10$ $GeV^2$ fit well the CLEO data  and  is in a good agreement with
theoretical predictions from the light-cone QCD sum rules (LCSR), offered in \cite{Khodjamirian:1997tk},
however, at larger virtualities strong disagreement takes place. Moreover, more precise recent  LCSR analysis \cite{Khodjamirian:2009ib, Mikhailov:2009kf} shows, that it is impossible to explain BaBar data on
$F_{\pi\gamma}(Q^2)$  at large $Q^2$ by use of usual (endpoint-suppressed) form of pion distribution
amplitude. This have led (quite unexpectedly) to the question of pQCD factorization validity.
Recently, there were proposed several approaches to explain such anomalous behaviour of $F_{\pi\gamma}(Q^2)$
\cite{Dorokhov:2009dg,Radyushkin:2009zg,Polyakov:2009je,Chernyak:2009dj,Dorokhov:2010bz}, in particular,
questioning pQCD factorization. At the same time, in \cite{Mikhailov:2010ud} the authors give some arguments
against the related flat-type pion distribution amplitude,  while in \cite{Roberts:2010rn} some doubts about
BaBar results analysis were expressed.

In this paper we study what can be learnt about the
meson-photon transition form factors from the anomaly sum rule
for the case of one virtual photon. This generalizes the usual
application of anomaly, providing the boundary condition in the
real photon limit only. Our (non-perturbative) QCD method does not imply the QCD
factorization and is valid even if the QCD factorization is broken.
It is shown, that using axial anomaly in the dispersive approach
we can get the exact relations between possible corrections to
lower states and continuum providing a possibility of relatively
large corrections to the lower states.

\section{Anomaly sum rule}
Following \cite{Horejsi:1994aj}, we briefly remind some results
of the dispersive approach to axial anomaly which are relevant
for this paper.
The VVA triangle graph amplitude
\be T_{\alpha \mu\nu}(k,q)=\int
d^4 x d^4 y e^{(ikx+iqy)} \langle 0|T\{ J^5_\alpha(0) J_\mu (x)
J_\nu(y) \}|0\rangle \ee
contains axial current $J^5_\alpha=(\bar{u}\gamma_5 \gamma_\alpha u -\bar{d}\gamma_5\gamma_\alpha d)$ and two
vector currents $J_\mu = ((2/3)\bar{u}\gamma_\mu u -(1/3)\bar{d}\gamma_\mu d)$;  $k,q$ are momenta of photons.
This amplitude can be presented as a tensor decomposition

\begin{eqnarray}
\label{eq1} \nonumber T_{\alpha \mu \nu} (k,q) & = & F_{1} \;
\varepsilon_{\alpha \mu \nu \rho} k^{\rho} + F_{2} \;
\varepsilon_{\alpha \mu \nu \rho} q^{\rho}
\\
& & + \; \; F_{3} \; q_{\nu} \varepsilon_{\alpha \mu \rho \sigma}
k^{\rho} q^{\sigma} + F_{4} \; q_{\nu} \varepsilon_{\alpha \mu
\rho \sigma} k^{\rho}
q^{\sigma}\\
\nonumber & & + \; \; F_{5} \; k_{\mu} \varepsilon_{\alpha \nu
\rho \sigma} k^{\rho} q^{\sigma} + F_{6} \; q_{\mu}
\varepsilon_{\alpha \nu \rho \sigma} k^{\rho} q^{\sigma} \;,
\end{eqnarray}
where the coefficients $F_{j} = F_{j}(k^{2}, q^{2}, p^{2}; m^{2})$, $p
= k+q$, $j = 1, \dots ,6$ are the corresponding Lorentz invariant
amplitudes (form factors). Note that these form factors
do not have kinematical singularities and are suitable for dispersive approach,
which we use to derive anomaly sum rule.

Symmetries of the amplitude $T_{\alpha \mu \nu}(k,q)$ impose the relations for the form factors $F_{j}(k,q)$.

Bose symmetry, i.e. $T_{\alpha \mu \nu} (k,q) = T_{\alpha \nu \mu} (q,k)$ leads to
\begin{eqnarray}
\label{eq2}
\nonumber
F_{1} (k,q) = - F_{2} (q,k), \\
F_{3} (k,q) = - F_{6} (q,k), \\
\nonumber
F_{4} (k,q) = - F_{5} (q,k).
\end{eqnarray}
Vector Ward identities for the  amplitude $T_{\alpha \mu \nu}(k,q)$
\begin{equation}
\label{eq3}
k^{\mu} T_{\alpha \mu \nu} = 0, \qquad q^{\nu} T_{\alpha \mu \nu}
= 0
\end{equation}
in terms of form factors read:

\begin{eqnarray}
\label{eq4}
F_{1} = k.q \, F_{3} + q^{2} \, F_{4} \;, \qquad
F_{2} = k^{2} \, F_{5} + k.q \, F_{6}\;.
\end{eqnarray}

Anomalous axial-vector Ward identity for
$T_{\alpha \mu \nu} (k,q)$  \cite{Bell:1969ts,Adler:1969gk}
\begin{equation}
\label{ward1} p^{\alpha} T_{\alpha \mu \nu} (k,q) = 2 m T_{\mu \nu}
(k,q) + \frac{1}{2\pi^{2}} \varepsilon_{\mu \nu \rho \sigma}
k^{\rho} q^{\sigma} \;
\end{equation}
in  terms of form factors can be rewritten as follows:
\begin{equation}
\label{eq12} F_{2} - F_{1} = 2 m G + \frac{1}{2\pi^{2}} \;,
\end{equation}
where $G$ is a form factor, related to the 2nd rank pseudotensor
$T_{\mu \nu}$ involved in the ``normal term'' on the r.h.s. of
(\ref{ward1}):
\begin{equation}
\label{eq10} T_{\mu \nu} (k,q) = G \; \varepsilon_{\mu \nu \rho
\sigma} k^{\rho} q^{\sigma} \;.
\end{equation}

Writing the unsubtracted dispersion relations for the form factors
one gets the finite subtraction for axial current divergence resulting
in the anomaly sum rule which for the kinematical configuration we are interested in ($k^{2} =
0$, $q^{2} \not= 0$) takes the form \cite{Horejsi:1994aj}:
\begin{equation}
\label{ASR} \int_{4m^{2}}^{\infty} A_{3a}(t;q^{2},m^{2}) dt =
\frac{1}{2\pi} \;,
\end{equation}
where \be A_{3a}=\frac{1}{2}Im (F_3-F_6). \ee It holds for an
arbitrary quark mass $m$ and for any $q^{2}$ in the considered
region. Another important property of the above relation is
absence of any $\alpha_s$ corrections to the integral \cite{Adler:1969er}. Moreover,
it is expected that it does not have any nonperturbative
corrections too ('t Hooft's principle).

\section{Transition form factors of mesons}

The form factor $F_{\pi\gamma}$ of the transition $\pi^0 \to
\gamma\gamma^*$ is defined from the matrix element:

\be \int d^{4}x e^{ikx} \langle \pi^0(p)|T\{J_\mu (x) J_\nu(0)
\}|0\rangle = \epsilon_{\mu\nu\rho\sigma}k^\rho q^\sigma
F_{\pi\gamma} \;, \ee
 where $k,q$ are momenta of virtual photons, $p=k+q$, and $J_\mu = ((2/3)\bar{u}\gamma_\mu u
 -(1/3)\bar{d}\gamma_\mu d)$ is the electromagnetic current of light quarks.

Three-point correlation function $T_{\alpha \mu\nu}(k,q)$ has
pion (pole at $p^2=m^2_\pi$) and higher states contributions:

\be T_{\alpha
\mu\nu}(k,q)=\frac{i\sqrt{2}f_\pi}{p^2-m_{\pi}^2}p_\alpha k^\rho
q^\sigma \epsilon_{\mu\nu\rho\sigma} F_{\pi\gamma}+ (higher\;
states) \;,
\ee where $f_\pi$ is a pion decay constant, which can be defined
as a coefficient in the projection of axial current $J^5_\alpha$ onto
one-pion state: \be \langle 0|J^5_\alpha(0) |\pi^0(p)\rangle=
i\sqrt{2}p_\alpha f_\pi \;. \ee The pion decay constant
$f_\pi=130.7$ MeV is experimentally  well determined from the
decay of charged pion $\pi^- \to \mu^- \nu$.
Using the kinematical identities
\begin{equation}
\delta_{\alpha\beta} \epsilon_{\sigma\mu\nu\tau}
-\delta_{\alpha\sigma} \epsilon_{\beta\mu\nu\tau}
+\delta_{\alpha\mu}\epsilon_{\beta\sigma\nu\tau}
-\delta_{\alpha\nu}\epsilon_{\beta\sigma\mu\nu}
+\delta_{\alpha\tau}\epsilon_{\beta\sigma\mu\nu}=0 \; ,
\end{equation}
we can single out the pion contribution to
$\frac{1}{2}(F_3-F_6)$. Then, the contribution of pion to $Im
(F_3-F_6)/2$ (imaginary part is taken w.r.t. $p^2$) is:

\be \label{pi} \frac{1}{2}Im (F_3-F_6) = \sqrt{2} f_\pi \pi
F_{\pi\gamma}(Q^2) \delta (s-m_\pi^2) \;, \ee
where $Q^2=-q^2$ \footnote{Note that in the case of the choice of the basis differing from (\ref{eq1}) for
AVV amplitude the results for transition formfactors do not change.}.

It is well known that at $Q^2=0$ the pion contribution saturates
anomaly sum rule (\ref{ASR}) \cite{Dolgov:1971ri} and
$F_{\pi\gamma}$ is known to be normalized by the
$\pi^0\rightarrow \gamma\gamma $ decay rate \cite{Adler:1969gk}:

\be F_{\pi\gamma}(0)=\frac{1}{2\sqrt{2}\pi^2 f_\pi}. \ee

On the other hand at $Q^2\neq0$, factorization approach to
perturbative quantum chromodynamics (pQCD) for exclusive process
in the leading order in the strong coupling constant predicts
\cite{Lepage:1980fj,Brodsky:1981rp}:

\begin{equation}\label{Fpqcd}
F_{\pi\gamma}(Q^{2})=\frac{\sqrt{2}f_\pi}{3Q^{2}}
\int_{0}^{1}dx\frac{\varphi_{\pi}\left( x,Q^2
\right)}{x}+\mathcal{O}(1/Q^4) \;,
\end{equation}
where $f_\pi=130.7$ MeV and  $\varphi_{\pi}(x)$ is a pion
distribution amplitude (DA). The pion DA depends on the
renormalization scale \cite{Efremov:1979qk,Lepage:1980fj} and at
large $Q^2$ asymptotically acquires a simple form
\cite{Efremov:1978rn} $\varphi_{\pi}^{\mathrm{asymp}}\left(
x\right)=6x\left( 1-x \right)$. This leads to asymptotic
behaviour for the pion form factor:

\begin{equation}\label{AsLargeQ}
F_{\pi\gamma}^{\mathrm{asymp}}(Q^{2})=\frac{\sqrt{2} f_{\pi}}{Q^{2}}+\mathcal{O}(1/Q^4) \;.%
\end{equation}

From (\ref{Fpqcd}) and (\ref{pi}) we get the contribution of pion
to anomaly sum rule (\ref{ASR}):

\be 2\pi f_\pi^2/Q^2 \;. \ee

We see, that at $Q^2\neq0$ anomaly sum rule (\ref{ASR}) cannot be
saturated by pion contribution \footnote{This may be compared with $Q^2$ dependence
\cite{Narison:1992fd} of pion-to-photons matrix element
emerging in the analysis of sum rule \cite{Narison:1992fd,Efremov:1989zx}
for photon spin structure function.} due to $1/Q^2$ behavior, so we
need to consider higher states. The higher mass pseudoscalar
states have the same behavior and suppressed by the factor
$m^2_\pi/m^2_{res}$ as follows from the PCAC (since $\partial_\mu
J^3_\mu$ should vanish in the chiral limit). The other
contributions are provided by axial mesons, the lightest of which
is the $a_1(1260)$ meson. In fact, the  contribution  of longitudinally
polarized  $a_1$ is given by the similar equation to
(\ref{AsLargeQ}) at large $Q^2$. indeed, the analysis of
axial current bilocal matrix elements of $a_1$ is completely similar for that \cite{Ball:1996tb}
of vector current matrix elements of $\rho$.  The
contribution of transversally polarized $a_1$ to (\ref{ASR})
decreases even faster. Actually, the same is  true for all the
higher axial mesons and mesons with higher spin. So we make an
important observation: for the case $Q^2\neq 0$ the anomaly relation
(\ref{ASR}) cannot be explained in terms of any finite number of
mesons due to the fact that all transition form factors are
decreasing functions. That is why we conclude that only
\textit{infinite} number of higher states can saturate anomaly
sum rule and therefore at $Q^2\neq0$ the \textit{axial anomaly is
a genuine collective effect of meson spectrum} in contrast with
the case of two real photons $Q^2=0$, where the anomaly sum rule
is saturated by pion contribution only. Let us note that this
conclusion does not depend on any choice of meson distribution amplitudes.

\section{Quark-hadron duality}
Now we proceed to particular analysis of anomaly sum rule
(\ref{ASR})using quark-hadron duality. In some sense the
discussion of the previous section is based on the quark-hadron
duality. In this section we will apply local quark-hadron duality
to anomaly sum rule (\ref{ASR}).
According to the quark-hadron duality, let us saturate the the
spectral density $A_{3a}$ by pion and continuum contributions:
\begin{equation} \label{spec}
A_{3a}\left(s,Q^2\right)= \sqrt{2}\pi f_{\pi}\delta(s-m_\pi^2)
F_{\pi \gamma}\left(Q^2\right)\ +
A^{QCD}_{3a}\theta(s-s_0),
\end{equation}
where continuum contribution $A^{QCD}_{3a} \theta(s-s_0)$ as
usually supposed to be equal to QCD calculated spectral function
$A_{3a}$, while $s_0$ is a continuum threshold.

Substituting (\ref{spec})  into (\ref{ASR}), we obtain the
anomaly sum rule in the following form:
\be \label{pi+cont1} \frac{1}{2\pi}=\sqrt{2}\pi f_\pi
F_{\pi\gamma}(Q^2)+ \int_{s_0}^{\infty}ds A^{QCD}_{3a} \;. \ee

One-loop PT calculation \cite{Gorsky:1987mu,Radyushkin:1996tb}
leads to a simple result for the spectral density function:
\begin{equation}\label{specQCD}
A^{QCD}_{3a}(s,Q^2) = \frac{1}{2\pi}{{Q^2}\over{(s+Q^2)^2}} \;,
\end{equation}

so we can rewrite the anomaly sum rule in the following form: \be
\label{pi+cont} \frac{1}{2\pi}=\sqrt{2}\pi f_\pi
F_{\pi\gamma}(Q^2)+
\frac{1}{2\pi}\int_{s_0}^{\infty}ds\frac{Q^2}{(s+Q^2)^2} \;, \ee

and finally the pion form factor is \be \label{qq}
F_{\pi\gamma}(Q^2)=\frac{1}{2\sqrt{2}\pi^2
f_\pi}\frac{s_0}{s_0+Q^2} \;, \ee where $s_0=0.7$ $GeV^2$ is a
continuum threshold.

This result coincides with the interpolation formula proposed by
S. Brodsky and G.P. Lepage \cite{Brodsky:1981rp} $(s_0=4\pi^2
f^2_\pi)$
\begin{equation}
F_{\pi\gamma}^{\mathrm{BL}}(Q^{2})=\frac{1}{2\sqrt{2}\pi^2
f_{\pi} }\frac{1}{1+Q^{2}/\left( 4\pi^{2}f_{\pi}^{2}\right)}
\;,\label{FpqcdBL}
\end{equation}
which was derived in the quark-hadron duality context by
A.V.~Radyushkin \cite{Radyushkin:1995pj}. The contact with exact
anomaly sum rule observed here allows to find the relations
between contributions of $\pi^0$ with higher states which may be
chosen  as $a_1$ and continuum. With account of these
contributions the anomaly sum rule (\ref{ASR}) can be rewritten
in the following form:

\be \label{pi+a1+cont} \frac{1}{2\pi}=\sqrt{2}\pi f_\pi
F_{\pi\gamma}(Q^2)+
I_{a_1}+\frac{1}{2\pi}\int_{s_1}^{\infty}ds\frac{Q^2}{(s+Q^2)^2}
\;, \ee where $I_{a_1}$ is a contribution of $a_1$ meson to sum
rule(which can be expressed in terms of $a_1$ form factors),
$s_1=2.5$ $GeV^2$ is a continuum threshold for this case.

Using the asymptotic formula for pion form factor (\ref{qq}) we
can estimate  the behavior of $I_{a_1}$ at large $Q^2$ as

\be I_{a_1}=\frac{1}{2\pi}Q^2\frac{s_1-s_0}{(s_1+Q^2)(s_0+Q^2)}
\;. \ee This equation can be treated as a good interpolation for
$a_1$ contribution with correct asymptotic behavior (large and
small $Q^2$).

The plot for contributions of pion, $a_1$ meson and continuum is shown in Fig.1.
The figure illustrates the anomaly collective effect: indeed, the
contribution of infinite number of higher resonances (continuum
contribution) dominates starting from relatively small $Q^2\simeq
1.5$ $GeV^2$.

\begin{figure}
  \includegraphics[width=0.6\textwidth]{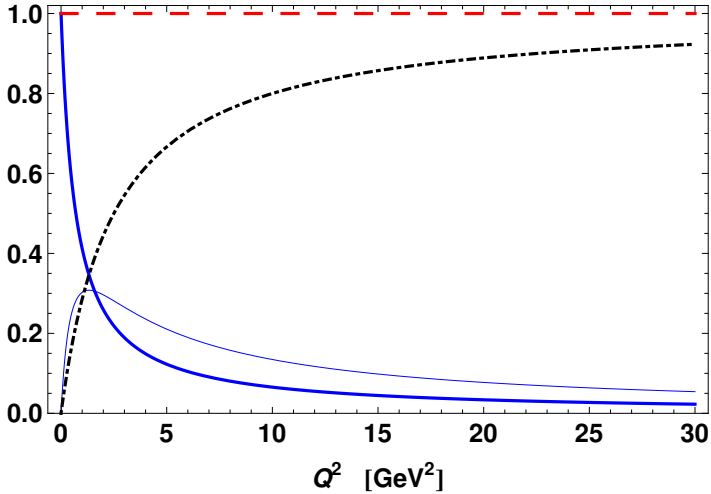}\\
  \caption{\textit{(Colour online).} Relative contributions of $\pi^0$ (thick blue curve), $a_1$ (thin blue
  curve) mesons (intervals of duality are $0.7$ $GeV^2$ and $1.8$ $GeV^2$ respectively), and continuum (
  dot-dashed black curve) (continuum threshold is $s_1=2.5$ $GeV^2$) to the anomaly sum rule (dashed red
  line).}\label{fig1}
\end{figure}

\section{Corrections' interplay and experimental data}
As we learned above, axial anomaly is a collective effect of
meson spectrum. That is why it is natural study the relation
between possible meson and continuum corrections  from anomaly
sum rule. Let us write out anomaly sum rule once more in the
following form:

\be \label{ASR1} \frac{1}{2\pi}=\int_{0}^{\infty}
A_{3a}(s;Q^{2})ds=I_\pi + I_{a_1} + I_{cont} \;, \ee where
continuum contribution $I_{cont}$ takes into account all other
higher mass axial and higher spin states.

As we already mentioned,  the anomaly sum rule (\ref{ASR}) is an
exact relation ($\int_{0}^{\infty} A_{3a}(s;Q^{2})ds$ does not
have any corrections). However, the continuum contribution
\be
I_{cont}=\int_{s_i}^{\infty} A_{3a}(s;Q^{2})ds  \ee
may have perturbative as well as power corrections.

Note that  the two-loop corrections to the whole triangle graph
were found to be zero \cite{Jegerlehner:2005fs} implying the zero corrections
to all spectral densities\footnote{Zero two loop corrections to spectral densities
were also found in the massive case
\cite{Pasechnik:2005ae}, although
later the inconsistency of this result with asymptotic mass expansion was pointed
out \cite{Melnikov:2006qb}.}.
To match this result with the earlier found non-zero corrections in the factorization approach
(see \cite{Mikhailov:2009kf} and references therein) one should be careful. When one is dealing with
the corrections to the form factor itself, only the corrections to the coefficient
function should be considered, while all other corrections
are absorbed to the definition of the distribution amplitude.
At the same time, when the factorization theorem is applied for calculations of the
large $Q^2$ asymptotics of VVA diagram, {\it all} the corrections should be taken into account.
To do so, one should add Eq. (3.11)  of \cite{Mikhailov:2009kf} with the projector to
the local axial current (proportional to asymptotic pion
distribution amplitude\footnote{We are indebted to S.V. Mikhailov for clarification of this point.})
and Eq. (B1) (coinciding with Eq. (1)
of the Erratum) of \cite{Bakulev:2001pa} and get a zero result, compatible with
\cite{Jegerlehner:2005fs}.

Therefore, the model of the corrections to continuum discussed
below should rather correspond to some non-perturbative corrections.
Let us first consider the contributions of local condensates. Naively, they should
strongly decrease with $Q^2$ compensating the mass dimension of gluon
(as quark one is suppressed even more)
condensate. However the 't Hooft's principle requires
(see \cite{Horejsi:1994aj}, Section 4)
the rapid decrease of the corrections with Borel parameter $M^2$ (related to $s$) so that
the power of $Q^2$ in the denominator may be not so large.
In reality the actual calculations do not satisfy
this property and the situation  may be
improved  by the use of non-local condensates
(see \cite{Horejsi:1994aj} and references therein). Another
possibility is other non-perturbative contributions,
like instanton-induced ones. So we assume the appearance of such
corrections in what follows modelling the corrections to continuum.

In order to preserve the sum
rule (\ref{ASR1}) (or in particular
(\ref{pi+cont}), (\ref{pi+a1+cont})) the corrections to continuum
contribution  should be exactly compensated by corrections to
lower states, in particular to pion. It turns out that this is a
rather uncommon situation: corrections to continuum are
compensated by the main terms of lower states which are of the
same order in $Q^2$ as continuum corrections.

To be more specific, let us consider the model
``$\pi^0$+continuum''.
From (\ref{pi+cont})  the main contributions of pion and
continuum read: \be I^0_\pi=\sqrt{2}\pi f_\pi
F^0_{\pi\gamma}(Q^2)= \frac{1}{2\pi}\frac{s_0}{s_0+Q^2}\;,\\
\ee
\be I^0_{cont}= \frac{1}{2\pi}\frac{Q^2}{s_0+Q^2}
\;. \ee

If the corrections to pion and continuum to ASR are $\delta I_\pi$ and
$\delta I_{cont}$ respectively \be  I_\pi=I^0_\pi+\delta I_\pi \;,
\ee
\be I_{cont}=I^0_{cont}+\delta I_{cont} \ee then, since $\delta
I_\pi=-\delta I_{cont}$, the ratio of relative corrections to
continuum and pion is \be \label{ratio}
 \left|\frac{\delta I_{cont}/I^0_{cont}}{\delta I_\pi/I^0_\pi} \right |=\frac{s_0}{Q^2} \;.
\ee
For instance, for $Q^2=20$ $GeV^2$, $s_0=0.7$ $GeV^2$ the ratio is
\be
 \left|\frac{\delta I_{cont}/I^0_{cont}}{\delta I_\pi/I^0_\pi} \right |\simeq 0.03 \;.
\ee
We see, that the relative correction to continuum is suppressed by factor $1/Q^2$
as compared to the correction to pion.
 To illustrate our conclusion, we assume the correction to
continuum at large $Q^2$ is $\delta I_{cont}=-c s_0
\frac{\ln(Q^2/s_0)+b}{Q^2}$. This correction preserves asymptotics
of continuum  contribution at large  $Q^2$. Contributions of pion
and continuum to ASR then have the following explicit forms: \be
\label{fit} I_{cont}= \frac{1}{2\pi}\frac{Q^2}{s_0+Q^2} - c
s_0\frac{\ln(Q^2/s_0)+b}{Q^2} \;, \ee

\be \label{fit2} I_\pi=
\frac{1}{2\pi}\frac{s_0}{s_0+Q^2}+ c s_0
\frac{\ln(Q^2/s_0)+b}{Q^2} \;. \ee
If this correction corresponds, as it was discussed above, to (non-local) gluon
condensate $\langle G^2 \rangle$ one may formally substitute the dimensional factor $c s_0$ by $c \langle G^2 \rangle /s_0$ stemming from $\langle G^2 \rangle /M^2$ for Borel transforms.

One can see, that the leading power correction to
continuum preserving
its asymptotics results in a substantial (of
the order of the main term $I^0_\pi$) contribution to the pion
state changing the pion form factor
asymptotics at large $Q^2$.

The experimental data on pion form factor behavior at large $Q^2$
allows us to get  estimation for the corrections to continuum. If
the  pion form factor expression (\ref{qq})  have matched the
experimental data, the continuum leading correction might be only
of order $1/Q^4$. However, the last BaBar data on pion transition
form factor \cite{Aubert:2009mc} manifests  large discrepancy  of
$F_{\pi\gamma}$ values at large $Q^2$ with the expected
asymptotic behaviour (\ref{qq}).

Relying on the BaBar data, we can fit  parameters $b,c$:
\be \label{fitN} b=-2.74,\;\;\;c=0.045.
\ee
The corresponding plot of combination $Q^2 F_{\pi\gamma}$ for the best-fit
parameters (\ref{fitN}) is shown in Fig. 2.

Basing on the model ``pion+continuum'' we can calculate the
relative correction to continuum contribution to ASR $\delta
I_{cont}/I^0_{cont}$ relying on different fits of
$F_{\pi\gamma}^*$: \be \left|\delta I_{cont}/I^0_{cont}\right |=
\sqrt{2}\pi f_\pi(F_{\pi\gamma}^*- F_{\pi\gamma}^0)/I^0_{cont}=
\frac{2\sqrt{2}\pi^2 f_\pi(s_0+Q^2)}{Q^2}F_{\pi\gamma}^*
-\frac{s_0}{Q^2}. \ee
In Fig. 3 the ratios $\delta I_{cont}/I^0_{cont}$ for our fit
(\ref{fit}, \ref{fitN}) and fit, obtained in recent paper
\cite{Radyushkin:2009zg} are shown. We see, that the correction to continuum is indeed small, even though
the BaBar data shows that the \textit{relative} correction to
pion contribution is large.

\begin{figure}
  \includegraphics[width=0.6\textwidth]{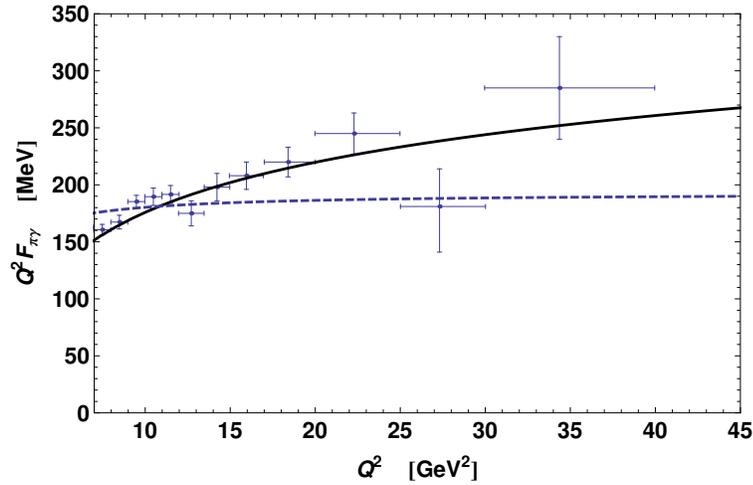} \\
  \caption{$Q^2 F_{\pi\gamma} (Q^2)$: BaBar data\cite{Aubert:2009mc}, pQCD calculated behaviour (\ref{qq})
  (dashed curve) and logarithmically enhanced behaviour due to small correction to continuum (\ref{fit2},
  \ref{fitN}) (solid curve).
  }
  \label{fig2}
\end{figure}

\begin{figure}
  \includegraphics[width=0.6\textwidth]{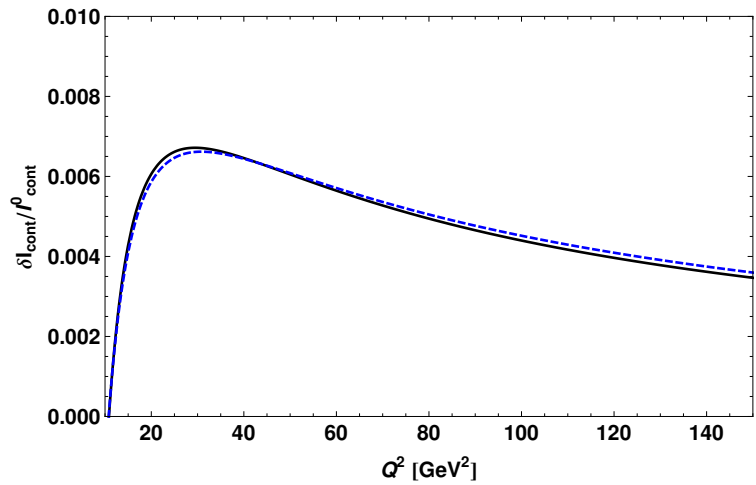} \\
  \caption{\textit{(Colour online).} Relative correction to continuum $\delta I_{cont}/I^0_{cont}(Q^2)$: our
  fit (\ref{fit}, \ref{fitN}) (solid curve) and fit, obtained in \cite{Radyushkin:2009zg} (dashed blue
  curve). }\label{fig3}
\end{figure}

Above we estimated the correction to continuum in the model
``$\pi^0$+continuum''.
However, one can consider more refined models
like   ``$\pi^0$ + $a_1$+continuum''.
This model may assume interplay
between corrections to three terms in the ASR.

Moreover, in the case of the small correction to continuum which currently
seems to be the most likely situation,
ASR will lead to the relation between the transition form factors of
pion and $a_1$ which may be further studied both theoretically and, most
important, experimentally.

\section{Discussion and Conclusions}
Dispersive approach to axial anomaly  proved to be a useful tool
for studying the  properties of meson spectrum. It is well known,
that when both photons are real ($Q^2=0$) the ASR saturates by
pion contribution only. However, when one of the photons is
virtual ($Q^2\neq 0$) we immediately get different situation: ASR
can be saturated only with a full meson spectrum (any finite
number of mesons cannot saturate the anomaly sum rule). So the
axial anomaly is a  collective effect of meson spectrum.

The anomaly  sum rule and quark-hadron duality in case of model
``$\pi^0$+continuum'' allows  to reproduce the well-known
Brodsky-Lepage interpolation formula for $F_{\pi\gamma}$. We
estimate  the contribution of $a_1$ meson to anomaly sum
rule in the model ``$\pi^0$+$a_1$+continuum''.

The exactness of the anomaly sum rule leads to the relation
between corrections to continuum and lower mass states
contributions. The last experimental data on pion transition form
factor $F_{\pi\gamma}$ at large $Q^2$  allows us to  estimate
the possible continuum correction.

One can also consider $\gamma^* \gamma \to \eta$ transition form
factor in the same way. Considering the similar ratio for the
relative contributions of $\eta$ meson and continuum as (\ref{ratio}) (with the
continuum threshold
\mbox{$s_0^\eta \sim 2.5$ GeV$^2$),}
we can estimate the
relative correction to $\eta$ to be several times smaller than
the one for $\pi^0$: \be
 \left|\frac{\delta I_{\eta}/I^0_{\eta}}{\delta I_\pi/I^0_\pi} \right |\simeq
 \frac{s_0^{\pi}}{s_0^{\eta}} \simeq 0.3.
\ee

We thank V.~M.~Braun, B.~L.~Ioffe, A. Khodjamirian, S.~V.~Mikhailov, S.~Narison  and
A.~V.~Radyushkin for useful discussions and elucidating comments. This  work
was supported in part by RFBR (Grants 09-02-00732, 09-02-01149, 08-02-01003),
by the funds from EC to the project
``Study of the Strong Interacting
Matter''  under contract No. R113-CT-2004-506078 and
by  CRDF  Project  RUP2-2961-MO-09.

\bibliographystyle{epj}
\bibliography{bibliography_General}
\end{document}